\documentclass[aps,prd,twocolumn,preprintnumbers,nofootinbib,superscriptaddress,longbibliography]{revtex4-1}

\usepackage{tensor}
\usepackage{amsmath,amssymb}
\usepackage{amsfonts}
\usepackage{dsfont}
\usepackage{slashed}
\usepackage{graphicx}
\usepackage{subcaption}
\usepackage{caption}
\usepackage{import}
\usepackage{appendix}
\usepackage{hyperref}
\usepackage{siunitx}
\usepackage{setspace}
\usepackage{makecell}
\usepackage{float}
\usepackage{chemformula}
\usepackage{mathtools}
\usepackage[clock]{ifsym}
\hypersetup{hidelinks}
\graphicspath{{img/}}

\newcommand{\vac}{|0\rangle}
\newcommand{\hatbf}[1]{\hat{\mathbf{#1}}}
\newcommand{\bra}[1]{\langle #1 |}
\newcommand{\ket}[1]{| #1 \rangle}
\newcommand{\sz}[1][\alpha]{S^z_#1}
\newcommand{\sx}[1][\alpha]{S^x_#1}
\newcommand{\sy}[1][\alpha]{S^y_#1}
\newcommand{\sP}[1][\alpha]{S^+_#1}
\newcommand{\sM}[1][\alpha]{S^-_#1}
\newcommand{\annihil}{a_\alpha}
\newcommand{\creation}{a^\dag_\alpha}

\makeatletter
\renewcommand*{\@fnsymbol}[1]{\ensuremath{\ifcase#1\or \downarrow\or \uparrow\or \ddagger\or
   \mathsection\or \mathparagraph\or \|\or **\or \dagger\dagger
   \or \ddagger\ddagger \else\@ctrerr\fi}}
\makeatother

\begin{document}

\title{Spin-dependent dark matter scattering in quasi-two-dimensional magnets}
\author{Giacomo Marocco}
\thanks{gmarocco@lbl.gov}

\affiliation{Physics Division, Lawrence Berkeley National Laboratory, Berkeley, CA 94720, USA}

\author{John Wheater} 
\thanks{john.wheater@physics.ox.ac.uk}

\affiliation{Rudolph Peierls Centre for Theoretical Physics, University of Oxford, Parks Road, Oxford OX1 3PU, United Kingdom}
\begin{abstract}
    We study the prospects of detecting dark matter coupled to the spin of the electron, such that it may scatter and excite magnons -- collective excitations of electronic spins. We show that materials exhibiting long-range magnetic order where the spins are coupled only along a plane may act as directional dark matter detectors. These quasi-2D materials possess anisotropic dispersion relations and structure functions which induce a sidereal modulation in the excitation rate. We calculate the expected signal rate for some candidate (anti)ferromagnets, demonstrating a possible route to the direct detection of spin-dependent dark matter in the keV to MeV mass range.
\end{abstract}

\maketitle

\section{Introduction}
The Universe is apparently filled with a non-baryonic fluid known as dark matter (DM), whose exact particle nature is unknown. The direct detection of DM is a long-standing aim of particle physics \cite{goodmanDetectabilityCertain1985}, the basic principle of which is to measure energy depositions associated with DM scattering or absorption events. Many nuclear recoil experiments having set stringent limits on DM heavier than the proton \cite{LZ:2022lsv, PandaX:2024qfu,PhysRevLett.121.111302}, where DM-nucleon interactions can impart significant energy. For lighter DM, the scattering kinematics hinders the sensitivity of such experiments to the coupling of DM with the Standard Model. Much theoretical \cite{Essig:2011nj, Essig:2012yx, Graham:2012su, Essig:2015cda} and experimental \cite{ SuperCDMS:2020ymb, Hochberg:2021yud, SENSEI:2023zdf, XENONCollaboration:2022kmb, PhysRevLett.129.161804, QROCODILE:2024zmg, Gao:2024irf} progress has been made in searching instead for energy deposition on electrons. 

When the DM mass is below $\sim \SI{10}{MeV}$, the momentum transfer is long enough that one may think of the scattering events as taking place between DM and emergent, collective excitations in the target. Traditionally, much of the focus has been on DM candidates that couple in the non-relativistic limit to the number densities of the species in the target. Such couplings mainly excite phonons, for instance in superfluid helium \cite{Schutz:2016tid, Knapen:2016cue, Acanfora:2019con}, crystals \cite{Knapen:2017ekk, Griffin:2018bjn, Mitridate:2023izi, Bloch:2024qqo}, semiconductors \cite{PhysRevD.95.023013}, and superconductors \cite{PhysRevD.94.015019}. 

An alternative possibility is that the coupling may instead be spin-dependent, e.g. DM may couple predominantly to the spin of an electron~\cite{Liu:2021avx, Catena:2024rym}. In this case, if the target exhibits long-range magnetic order, the scattering can excite collective spin-wave excitations known as magnons. This effect has been previously studied~\cite{trickleDetectingLight2020, trickleEffectiveFieldTheory2020a, Esposito:2022bnu}, demonstrating that the magnon excitation rate may be much larger than the accompanying phonon rate.

Scattering in antiferromagnets (AFMs), in particular, has been shown to be optimal, owing to the favourable kinematics associated with the magnon's linear dispersion relation~\cite{Esposito:2022bnu}. Detection of these magnons thus seems a promising avenue for light DM detection. Previous work has operated in the effective field theory regime, restricting the range of DM masses that may be accurately predicted to $m_\chi \lesssim \SI{50}{keV}$. Here, we extend calculation of scattering in AFMs to the higher mass regime where the field theory breaks down, taking particular care to account for magnetic form factors that arise in the matching of microscopic electron interaction to the collective magnon description. 

We also extend existing calculations to a class of magnets that have intrinsic anisotropies, which allows for \textit{directional} direct detection. The DM excitation rate will exhibit a sidereal modulation as the Earth rotates through the DM wind, allowing for an additional experimental handle on the signal. Some ideas to this end for sub-GeV DM have been explored in the literature~\cite{Coskuner:2021qxo, Boyd:2022tcn}, for instance using polar materials \cite{Griffin:2018bjn}, defect production \cite{Budnik:2017sbu}, semiconductors \cite{Kadribasic:2017obi}, as well as intrinsically two-dimensional targets \cite{Hochberg:2016ntt}. 

In this paper, we consider  a method of directional detection of light DM candidates coupling to electronic spins. In particular, we demonstrate that magnon creation in quasi-1D and -2D (anti)ferromagnets are potentially sensitive to DM with mass in the 1 keV to 10 MeV range, with $\mathcal{O}(10\%)$ daily modulation. This modulation is a result of two effects: the intrinsically anisotropic dispersion relation of magnons in such materials, as well as the anisotropy of the magnetic spin's wavefunction at small (ionic) scales. In deriving this latter effect, we extend previous calculations of DM-magnon interactions \cite{trickleEffectiveFieldTheory2020a} to include a target-dependent magnetic form factor.

\textit{Conventions}: We take $i,j=1,2,3$ to be spatial indices; $a,b$ label electrons; $\alpha$ labels lattice sites. 
\section{Magnon excitation }

In order to assess the suitability of a target material for directional dark matter detection, we wish to evaluate the magnon excitation rate $\Gamma$ in the presence of a background DM field. If the DM is weakly coupled to electrons, we may evaluate this rate using linear response theory such that it factorises and is given by
\begin{equation}
\Gamma = \langle \int d^3 \mathbf{q}\,  V_{ij} \cdot S^{ij} \rangle_{\mathbf{v}_\chi},
\label{eqn:rateToy}
\end{equation}
where $V_{ij}$ is the scattering potential governed by the particular DM model in question, $S_{ij}$ is the spin-structure function that encodes the target response, the integral is over the momentum transfer $\mathbf{q}$, while $\langle \cdot \rangle _{\mathbf{v}_\chi}$ indicates an average over the incoming dark matter velocity distribution $f(\mathbf{v}_\chi)$.  The matrix $S_{ij}$ is the main object of interest for DM detection, which we will study in the rest of this section.

Our strategy will be as follows. First, we calculate $S_{ij}$ at the level of underlying interaction, i.e in terms of electrons and their spin. We then match this to the relevant IR modes, which in this case are those of lattice of effective spins. We calculate the structure function with this effective lattice theory, and explore its properties in (anti)ferromagnetic materials.

\subsection{Factorising the rate}

Let us begin by giving the exact form of the excitation rate given by Eq. \eqref{eqn:rateToy}. As stated, we are interested in coupling solely to the electron's spin.  Concretely, we focus on an interaction Hamiltonian \cite{fitzpatrickEffectiveField2013, trickleEffectiveFieldTheory2020a}
\begin{equation}
H_\mathrm{int} = \sum_a \mathbf{V}(\hat{\mathbf{r}}_\chi-\hat{\mathbf{r}}_a) \cdot \hat{\mathbf{S}}_{\mathrm{e},a},
\end{equation} 
where the sum is over electrons indexed by $a$ and the potential $\mathbf{V}$ depends only on the difference between the DM position $\mathbf{r}_\chi$ and the electron position $\mathbf{r}_a$. Within the Born approximation, the excitation rate is simple to derive. The detector excitation rate from its ground state $\vac$ to some final state $\lvert f \rangle$ reads
\begin{align}
\begin{split}
\Gamma(\mathbf{v_\chi}) &= \int \frac{d^3 \mathbf{q}}{(2\pi)^3 V_\mathrm{T}} \sum_{ij}  V_i(\mathbf{q}) V_j^\dagger(\mathbf{q}) \\
&\times \sum_f \langle f \rvert \hat{S}_{\mathrm{e},i}(\mathbf{q}) \lvert 0 \rangle \langle 0 \rvert \hat{S}_{\mathrm{e},j}^\dagger(\mathbf{q}) \lvert f \rangle \cdot 2\pi \delta(\omega_f - \omega_\mathbf{q}),
\label{eq:rate}
\end{split}
\end{align} 
where $\omega_\mathbf{q} = \mathbf{q} \cdot \mathbf{v}_\chi -q^2/2m_\chi$ is the energy transfer associated with a 3-momentum transfer $\mathbf{q}$, $V_\mathrm{T}$ is the target volume, $V_i(\mathbf{q})$ is the Fourier transform of $\mathbf{V}(\mathbf{x})$, and
\begin{equation}
\hat{\mathbf{S}}_\mathrm{e}(\mathbf{q}) = \sum_a e^{i \mathbf{q} \cdot \mathbf{x}_a } \hat{\mathbf{S}}_{\mathrm{e},a}
\end{equation}
is a momentum-weighted of electron spins. 
As previously noted, the detector-specific, many-body physics conveniently factorises and sits inside the spin matrix element, which make up the structure function
\begin{align}
    S_{ij} = \frac{1}{V_T} \sum_f \langle f \rvert \hat{S}_{\mathrm{e},i}(\mathbf{q}) \lvert 0 \rangle \langle 0 \rvert \hat{S}_{\mathrm{e},j}^\dagger(\mathbf{q}) \lvert f \rangle \cdot 2\pi \delta(\omega_f - \omega_\mathbf{q}).
\end{align}


\subsection{Matching to the lattice}
We have seen from Eq. \eqref{eq:rate} that the basic object of interest is $\langle f | \hat{\mathbf{S}}(\mathbf{q}) \vac$. We wish to match this to a lattice theory in which the basic degrees of freedom are some effective lattice spins
\begin{equation}
\hat{\mathbf{S}}_\alpha = \sum_{a \in \alpha} \hat{\mathbf{S}}_{\mathrm{e},a},
\label{eqn:totalSpin}
\end{equation} 
where $\alpha$ labels the lattice site and the sum is over all the electron spins at a site. For a crystal target, let us first then begin by calculating the matrix element for a single ion\footnote{Here, we follow the magnetic scattering convention that ion refers to a single magnetic atom.} to go from some initial lattice state $| \psi_i \rangle = |S_\alpha, S_z, \beta \rangle $ to some final lattice state $|\psi_f \rangle =  |S_\alpha', S_z', \beta' \rangle$
\begin{equation}
\mathcal{M}_{0f} = \sum_{a \in \alpha} \langle \psi_i | e^{i \mathbf{q} \cdot \hat{\mathbf{x}}_a} \hat{\mathbf{S}}_{\mathrm{e},a}  | \psi_f \rangle,
\end{equation}
where $S_\alpha$ labels the total spin, $S_z$ is the magnetic quantum number, and $\beta$ denotes the other quantum numbers of the site, including the electronic spatial wavefunctions (we have suppressed the lattice indices on all these quantities to avoid clutter). 

The evaluation of such matrix elements follows a standard procedure \cite{formFactors}. Assume first that the lattice site has some definite position $\mathbf{X}_\alpha$ that changes adiabatically upon scattering, and write the electron's position in terms of its relative position with respect to this $\mathbf{x}_a = \mathbf{X}_\alpha + \mathbf{u}_a$. Then, we require knowledge of
\begin{equation}
\mathcal{M}_{0f} = e^{i \mathbf{q} \cdot \hatbf{X}_\alpha}  \bra{\psi_i} \sum_a e^{i \mathbf{q} \cdot \hatbf{u}_a} \hatbf{S}_a \ket{\psi_f}.
\end{equation}
If $S_\alpha = S_\alpha'$, the total spin at the site is unchanged and we may make use of the Wigner-Eckart projection theorem, which states that for any SO(3)-vector operator $\hatbf{W}$, its matrix elements obey
\begin{equation}
\bra{S_\alpha, S_z, \beta} \hatbf{W} \ket{ S_\alpha, S'_z, \beta'} = \frac{\bra{S_\alpha, \beta} \hatbf{W}\cdot \hatbf{S} \ket{S_\alpha, \beta'} }{S_\alpha(S_\alpha+1)} \times \langle \hatbf{S}_\alpha \rangle,
\end{equation}
where $\langle \hatbf{S}_\alpha \rangle \equiv \bra{ S_\alpha, S_z, \beta} \hatbf{S}_\alpha \ket{S_\alpha, S'_z, \beta'}$.
Applying this to our expression with $\hatbf{W} = \sum_a e^{i \mathbf{q} \cdot \hatbf{u}_a} \hatbf{S}_a$, we find 
\begin{equation}
\bra{\psi_i} \sum_{a \in \alpha} e^{i \mathbf{q} \cdot \hatbf{u}_a} \hatbf{S}_a \ket{\psi_f} = f_{\alpha}(\mathbf{q}) \bra{\psi_i} \hatbf{S}_\alpha \ket{\psi_f},
\end{equation}
where $f_{\alpha}(\mathbf{q})$ is the ``magnetic form factor" and the remain expectation value is of the total spin at the site as defined in Eq. \eqref{eqn:totalSpin}.

\begin{figure}[hb!]
\centering
\includegraphics[width=\linewidth]{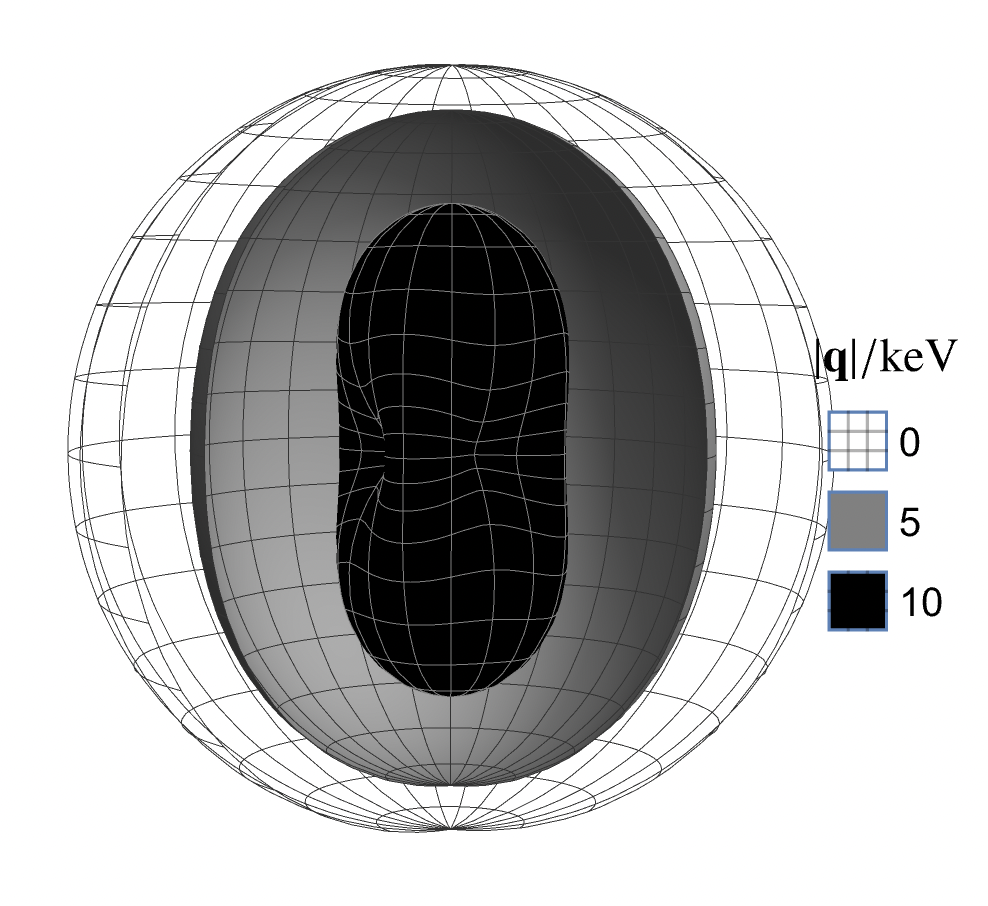}
\caption{The magnetic form factor $f(\mathbf{q})$ for an electron in the $3d_{x^2-y^2}$ orbital of a Cu$^{2+}$ ion, plotted over the sphere for three choices of $\lvert \mathbf{q} \rvert$; the radial coordinate of the surface denotes the magnitude of $f$. At large momentum transfer, the form factor deviates from $f(0) = 1$, with pronounced anisotropy in addition to damping.}
\label{fig:anistropicForm}
\end{figure}

The form factor, for the spin-only scattering considered here, is proportional to the electron charge density and is in general anisotropic, which will have important consequences for the daily modulation of the signal, as we will demonstrate. Its evaluation requires the electron wavefunctions, which have fortunately been calculated for ions of interest in the neutron scattering literature.  These form factors are tabulated, and we give the explicit form in App. \ref{app:formFactor} for the magnetic Cu$^{2+}$ ions that we study in this paper, which is plotted graphically in Fig. \ref{fig:anistropicForm}. In particular, we see that at large momentum transfer $|\mathbf{q}| \approx \SI{10}{keV}$ the scattering rate is heavily damped by the form factor, leading to a damping in the scattering rate by approximately two orders of magnitude.

It is important to recognise that this matching procedure was target specific, as we required knowledge of the magnetic ion and its electron's wavefunction. However, in the long wavelength limit, 
\begin{equation}
e^{i \mathbf{q} \cdot \hatbf{u}_a} \simeq 1
\end{equation}
and the matching to $\hatbf{S}_\alpha$ proceeds in a material-independent manner. This approximation is accurate to within $\mathcal{O}(10\%)$ for $q \lesssim \SI{10}{nm^{-1}}$, which is the characteristic momentum for $m_\chi \lesssim \SI{2}{MeV}$. The above multipolar expansion of the form factor, however, extends well into the ionic radius. 

%

To recap, we have shown that the dynamics of individual electrons may be matched to the dynamics of an effective lattice spin at wavelengths that are long with respect to the ionic size by introducing a magnetic form factor $f_\alpha$. The total matrix element is then a sum over lattice sites of this matched matrix element. Restricting ourselves to systems where all the lattice sites are equivalent, $f_\alpha = f$, the relevant matrix element may be written
\begin{equation}
\mathcal{M}_{0f}  = f(\mathbf{q}) \bra{0} \sum_\alpha e^{i \mathbf{q}\cdot \hatbf{X}_\alpha} \hatbf{S}_\alpha \ket{f},
\label{eqn:spinCorrelator}
\end{equation}
where the relevant operators act at level of the effective lattice theory.

\subsection{Spin-spin correlations and magnons}
With a lattice spin correlation function in hand, we now turn to its evaluation. We will make use of some standard results in the theory of magnetic excitations (for a brief review of magnets and magnons, see appendix \ref{app:magnons}). 

Consider the Heisenberg magnet, described by the Hamiltonian
\begin{equation}
\mathcal{H} = \sum_{\langle \alpha \beta \rangle} J_{\alpha \beta} \mathbf{S}_\alpha \cdot \mathbf{S}_\beta,
\end{equation}
where the sum is over nearest-neighbour sites $\langle \alpha \beta \rangle$ that are coupled with a strength $J_{\alpha \beta}$. The behaviour of the system depends strongly on the sign of $J_{\alpha \beta}$: if negative, the system is a FM, otherwise it is an AFM. We will discuss on the FM case.

\begin{figure*}[ht!]
\centering
\begin{subfigure}{.4\textwidth}
  \centering \includegraphics[width=\linewidth]{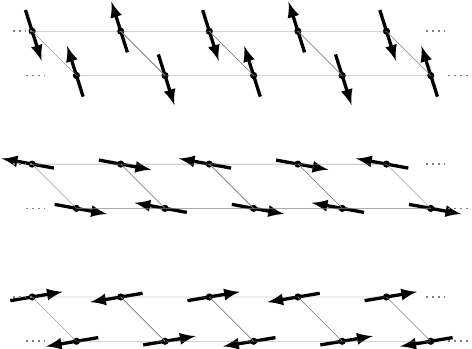}
  \caption{}
  \label{fig:spins}
\end{subfigure}
\begin{subfigure}{.43\textwidth}
  \centering \includegraphics[width=\linewidth]{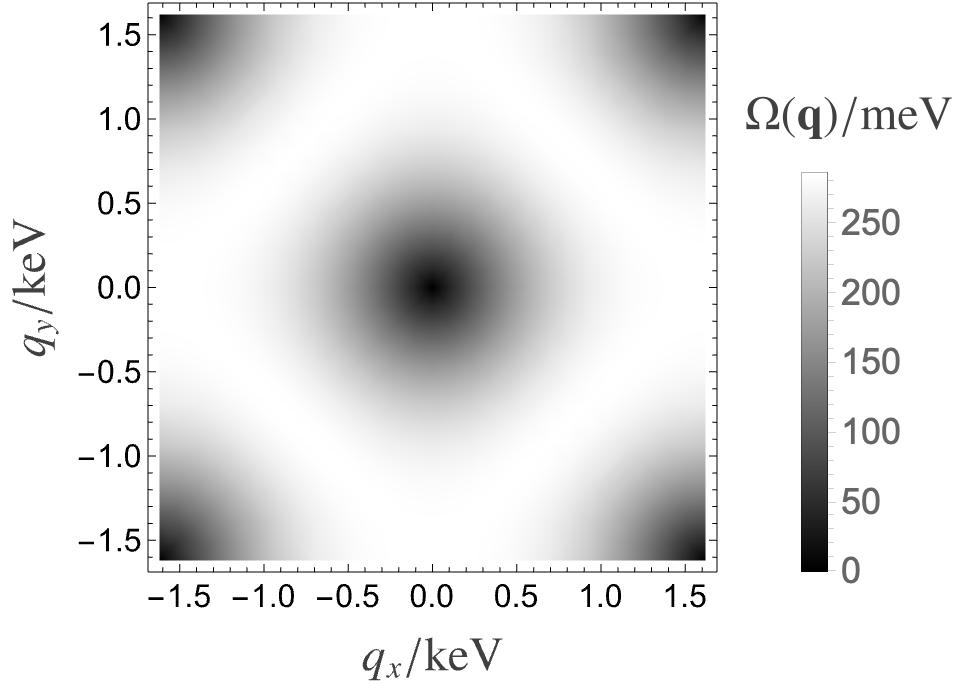}
  \caption{}
  \label{fig:dispersion}
\end{subfigure}
\caption{A square lattice, quasi-2d antiferromagnet has spins which are only coupled along certain planes, as depicted in Panel (a). The coupling between the spins (in the $x-y$ plane) gives rise to a non-trivial dispersion relation $\Omega(\mathbf{q})$ along the $q_x$ and $q_y$ directions, shown in Panel (b) for \ch{La2CuO4}. The dispersion relation is flat in the direction perpendicular to the planes (the $q_z$ direction), since  rotating spins between uncoupled planes requires no energy, introducing directionality into the system.}
\label{fig:anisotropicDispersion}
\end{figure*}

\subsubsection{Ferromagnetic magnons}
The low-lying excitations above the FM ground state are the single-magnon states, indexed by a crystal momentum $\ket{\mathbf{q}}$. The dynamics of single magnons are described in Appendix \ref{app:magnons}. At low energies, these modes dominate the system's dynamics, and the spin-spin correlation function may be approximated within the framework of linear spin-wave theory as
\begin{equation}
\begin{split}
\tilde{S}_{ij}(\omega, \mathbf{q}) &\equiv \frac{1}{V_\mathrm{T}} \sum_f \langle f \rvert \hat{S}_i(q) \lvert 0 \rangle \langle 0 \rvert \hat{S}_j^\dagger(\mathbf{q}) \lvert f \rangle  \delta(\omega_f(\mathbf{q}) - \omega) \\
\tilde{S}_{xx} &= \tilde{S}_{yy} = n_S\, \delta(\omega - \Omega(\mathbf{q})),
\end{split}
\label{eqn:spinspin}
\end{equation}
where $\omega_f$ is the energy of the final state, which is equal to the magnon's energy $\Omega$ in this regime and $n_S$ is the spin density of the system. At higher energies, there are more terms appearing in the sum over final states, but we may take this to be a lower bound on the size of the correlator.

The candidate FM we consider is \ch{K2CuF4}, a quasi-2d FM \cite{hirakawaNeutronScattering1983, ExperimentalDetermination1976a, yamadaMagneticProperties1972}. It has a square-lattice crystal structure (in the $x-y$ plane, say), where neighbouring sites within the square plane are coupled isotropically with strength $J_{\alpha \beta} = - J$, while the spins along the orthogonal plane are uncoupled\footnote{At long wavelengths, the inter-plane coupling, as well as anisotropic intra-plane couplings, are significant. Experimental observations indicate that only for $ q \gg 0.03 \,a^{-1}$ is the quasi-2$d$, isotropic, Heisenberg magnet description appropriate \cite{hirakawaNeutronScattering1983}. Since we focus on the regime $\omega \gtrsim$ meV, this approximation is valid.}. The spin can nonetheless point in any direction, not just in the magnetised plane, in contrast to so-called ``easy-plane" magnets. The resulting dispersion relation for this system is 
\begin{equation} 
\Omega(\mathbf{q}) = 4 J s \left( \sin^2 \frac{q_x a}{2} + \sin^2 \frac{q_y a}{2} \right);
\end{equation}
the relevant parameters are $a = \SI{4.125}{\AA} \simeq \SI{8.70}{keV^{-1}}$, $Js = \SI{1.0}{meV}$, and the perpendicular direction repeats with size $c = \SI{12.669}{\AA}$. Note that the dispersion relation is flat along the $z$-direction. This has important implications for DM direct detection, since at times of the day where the DM velocity is predominantly aligned with the flat direction, there will be less energy deposited in the target. This will cause a significant daily modulation, as we shall see.

\subsubsection{Anti-ferromagnetic magnons}
\begin{table*}[t]
\begin{tabular}{l|c|c|c|c|c|c|c| c} 
 Material & \; Type \; & \; $|J|$/\SI{}{meV} \; & \; $S$ \; & \; $a$/\SI{}{\angstrom} \; & \; $b$/\SI{}{\angstrom} \; & \; $c$/\SI{}{\angstrom} \; &  \; $\rho_\mathrm{T}$/\SI{}{g.cm^{-3}} \; & \; $n_S$/\SI{10^{21}}{cm^{-3}}   \\ 
 \hline
 \ch{La2CuO4} \cite{coldea2001spin, La2CuO4} & AFM  & 143 & 1/2 & \textbf{3.8} & \textbf{3.8} & 13.1 & 6.98 & 2.6 \\
 \ch{K2CuF4} \cite{yamadaMagneticProperties1972, hidaka1983x} & FM & 2 & 1/2 & \textbf{4.2} & \textbf{4.2} & 12.8 & 3.18 & 2.2  \\
\end{tabular}
\caption{The magnetically-ordered materials which we consider, along with their relavant properties. The type denotes whether the material is a ferromagnet (FM) or antiferromagnet (AFM). $|J|$ gives the strength of this coupling that enters into the Heisenberg magnet Hamiltonian. $S$ is the size of the spin. $a$, $b$, and $c$ give the lattice spacings, with directions along which spins are coupled in \textbf{bold}. $\rho_\mathrm{T}$ is the material density, and $n_S$ is the spin density derived from the previous quantities. }
\end{table*}

We now consider materials with antiferromagnetic order. In the Neel regime, where we may again carry out a semi-classical expansion, there exist two degenerate magnons, as described in App.~\ref{app:AFMs}. For square-lattice AFMs, whose spins are again coupled along the $x-y$ plane, the dispersion relation is 
\begin{align}
   \Omega(\mathbf{q}) = 4 J S \sqrt{1 - \frac{\left( \cos(q_x a) + \cos(q_y a)\right)^2}{4}},
\end{align}
which we plot in Fig~\ref{fig:anisotropicDispersion}b).
As with FMs, we see that the existence of flat directions in momenta. Furthermore, in the long-wavelength limit, the dispersion relation reduces to 
\begin{align}
    \Omega(\mathbf{q}) &\approx 4 J S q_\perp a \\
    &\approx \num{5\times 10^{-4}} \, q_\perp  \left( \frac{JS}{\SI{50}{meV}}\right) \left(\frac{a}{ \SI{5}{\angstrom}}\right)
\end{align}
where $q_\perp \equiv \sqrt{q_x^2 + q_y^2}$. This linear dispersion relation can allow for large density of states at small momenta than in an FM, which is favourable for DM scattering, in addition to having better kinematic matching to DM~\cite{Esposito:2022bnu}. Furthermore, we see that for characteristic momentum $q_\perp \geq \SI{10}{eV}$, corresponding to a dark matter mass $m_\chi \approx q_\chi/v_\chi \geq \SI{10}{keV}$, the magnon will have multi-meV energy.

The spin structure function for such a quasi-2d AFM, when summed over the two degenerate magnon modes, is
\begin{equation}
    \begin{split}
        \tilde{S}_{xx}(\mathbf{q}) &= \tilde{S}_{yy}(\mathbf{q}) = n_S \, \sqrt{\frac{2 - \cos q_x a - \cos q_y a}{2 + \cos q_x a + \cos q_y a}} \delta(\omega - \Omega(\mathbf{q})).
    \end{split} 
\end{equation}
Here we see that the AFM scattering function itself is anisotropic, which can lead to an enhanced anistropy. This features a divergence in the limit $q_x = q_y \to a \pi$, where the magnon energy vanishes at finite momentum transfer. The finite energy threshold of an experiment acts as a regulator in our case. We note that the singularity occurs at the edge of the Brillouin zone -- such an effect is invisible in the magnon EFT, since it occurs at the cut-off.
\section{Dark matter scattering}

The magnons that we have described in the previous section may be created via scattering of DM with the magnetic target. 

\subsection{DM models}

Although our results apply to any DM model that couples to electronic spin $\mathbf{S}_e$, for concreteness we will consider a model of fermionic dark matter $\chi$ coupling to the Standard Model photon through a magnetic dipole moment operator:
\begin{equation}
    \mathcal{L}_\mathrm{int} = e_V V_\mu \, \bar{e} \gamma^\mu e + \frac{1}{2} \mu_\chi \bar{\chi} \sigma^{\mu\nu} \chi V_{\mu\nu},
\end{equation}
Past constraints on this model have been derived, some of which assume the dark state constitutes the entirety of DM \cite{Pospelov:2000bq, Sigurdson:2004zp, Banks:2010eh, Kopp:2014tsa,Ibarra:2015fqa, Kavanagh:2018xeh}, while others are more model-independent \cite{L3:1999oof, Fortin:2011hv, Chu:2020ysb, Marocco:2020dqu}. As we will demonstrate, DM not excluded by any of these prior searches can induce possible measurable magnon excitation rates.

In App. \ref{app:MDM}, we derive the DM interaction potential resulting from this interaction via a massless mediator as 
\begin{align}
    V_{ij}(\mathbf{q}) = 2 \Big(\frac{e_V \mu_\chi}{m_e}\Big)^2 (\delta_{ij} - \hat{q}_i \hat{q}_j),
\end{align}
where $\hat{q}_i$ are unit vectors in the direction of momentum transfer. We see that this interaction projects onto the spin structure function perpendicular to the momentum transfer, and so there is an intrinsic directionality. However, this may not be the case for other interactions, for instance those of a pseudoscalar mediator. Furthermore, this directionality is washed out by the presence of multiple domains, as shown in App. \ref{app:domains}.

We also consider the non-relativistic, standard spin-dependent interaction potential
\begin{align}
    V(\mathbf{q}) = c \, \mathbf{S}_\chi \cdot \mathbf{S}_\mathrm{e},
\end{align}between our dark fermion $\chi$ and an electron, where $c$ is some coupling constant. This could arise, for instance, from integrating out the heavy axial-vector gauge boson of App.~\ref{app:axialVector}, in which case $c = 4 g_\chi g_e/m_V^2$, although we choose to remain agnostic to the UV completion. The scattering potential in this case is 
\begin{align}
    V_{ij}(\mathbf{q}) = |c|^2 \delta_{ij},
    \label{eqn:SDpotential}
\end{align}
which is see is independent of $\mathbf{q}$, and proportional to the identity.

Since the scattering potential Eq.~\eqref{eqn:SDpotential} is completely isotropic, it can introduce no daily modulation in an isotropic material. In particular, breaking of isotropy by the aligned spins in a single-domain magnet is not necessarily enough, as the potential picks out the trace of the spin structure function $S_{ij}$, which can be isotropic in this case even if the individual diagonal elements are not, as seen in the previous section. This necessitates using targets with intrinsic anisotropies, such as the quasi-low-dimensional materials we consider.

\subsection{Daily modulation of signal}

The anisotropy of the target is sensitive to the rotation of the Earth through the DM wind. Recall that we work in a co-ordinate system in which the crystal magnetisation plane is the $x-y$-plane. Then, aligning the $x$-axis with the Earth's velocity direction at $t=0$, the Earth's velocity may be written \cite{Coskuner:2019odd}
\begin{equation}
\mathbf{v}_e(t) = \hat{R}_x(\beta) \cdot \begin{pmatrix}
\cos^2 \Theta+ \sin^2 \Theta \cos \phi (t) \\ \sin \Theta \sin \phi (t) \\ \sin \Theta \cos \Theta (\cos \phi (t) -1)  
\end{pmatrix},
\end{equation}
where $\hat{R}_x(\beta)$ is a rotation around the $x$-axis through an angle $\beta$, $\Theta = 43^{\circ}$ is the angle between the Earth's axis of rotation and the velocity of the solar system in the galactic frame, and $\phi(t) = 2\pi t / T$ is the phase of the Earth's rotation, with $T = 24$ hours. We choose $\beta = 90^\circ$ so that the component of the velocity in the magnetisation plane oscillates with a period of half a day.

How does this oscillating velocity affect the energy of the excited magnon modes? We see that for small masses, the emergent $O(2)$ symmetry of the dispersion relation at long wavelengths causes the energy to oscillate in phase with the component of the Earth's velocity in the magnetic plane.

\begin{figure}[ht!]
\centering
\includegraphics[width=\linewidth]{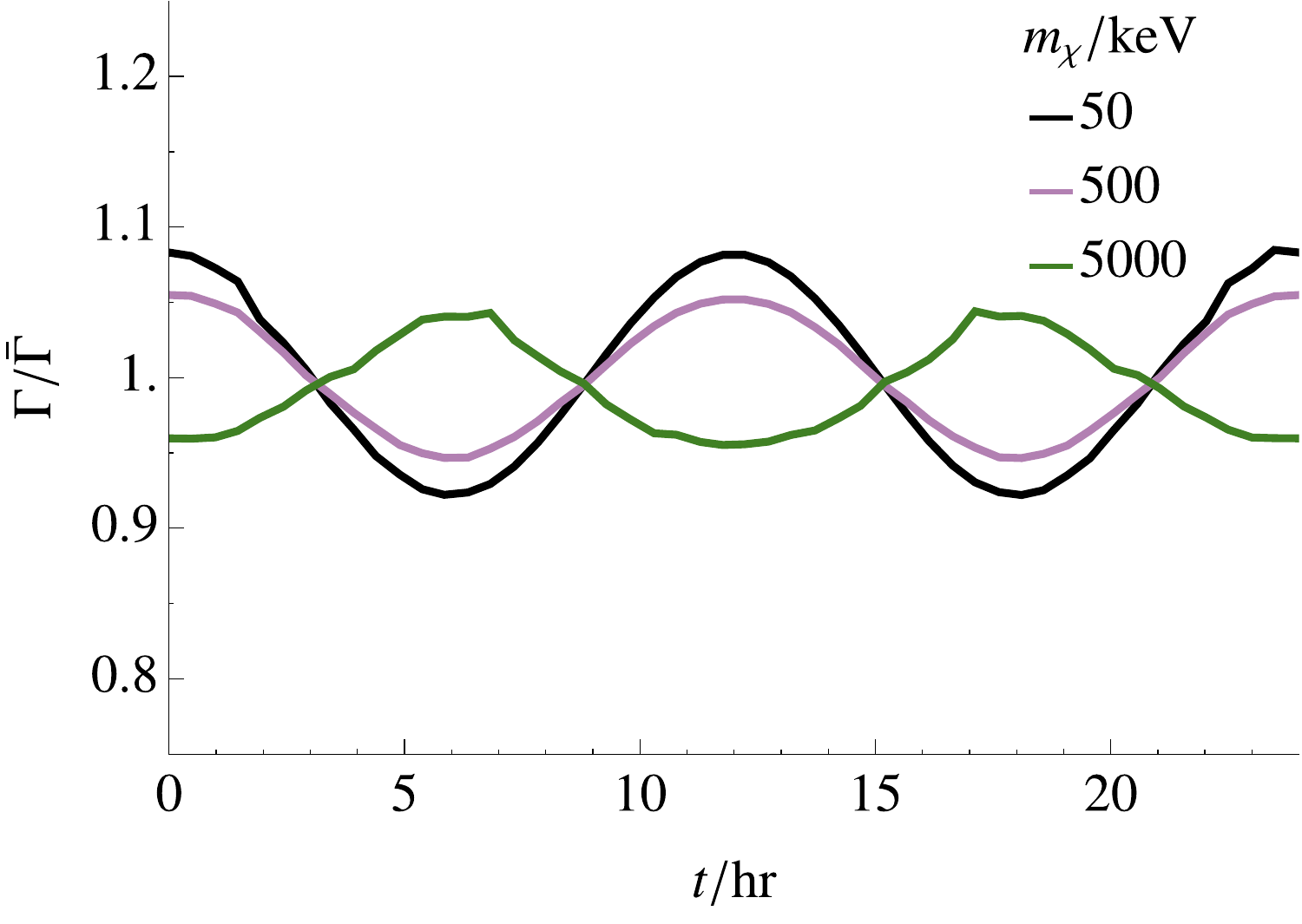}
\caption{The expected signal rate $\Gamma$ over a day for a \ch{La2CuO4} target, normalised to its mean value $\bar{\Gamma}$. The rate exhibits modulations of a few to ten percent over a range of DM masses $m_\chi$.}
\label{fig:dailyModulation}
\end{figure}

In Fig. \ref{fig:dailyModulation}, we plot the expected signal rate for the standard SD interaction over the course of a day, for a selection of DM masses. The absolute magnitude of the oscillation may be large -- as big as 10\%, even if the interaction potential is isotropic. 

This modulation can provide an additional handle on discriminating a signal against background. Assuming that most of the signal power is in the lowest harmonic, which oscillates with a period of a 12 hours (not 24, due to an approximate parity symmetry), the appropriate measure of this effect is given by the difference of the average rate between 6 hour bins
\begin{equation}
f_2(t_0) = \frac{1}{T \langle \Gamma \rangle_T} \left( \int_{t_0}^{t_0+\SI{6}{hr}} \Gamma(t)\, dt - \int_{t_0-\SI{6}{hr}}^{t_0}\Gamma(t) dt  \right),
\label{eqn:asymmetry}
\end{equation}
as introduced in \cite{blancoDarkMatter2021}. In this expression, $T = \SI{24}{hr}$, $\langle \Gamma \rangle_T$ is the time-averaged rate, while $t_0$ sets the phase of the oscillations. In terms of this measure, the statistical significance of an oscillating signal is given by
\begin{equation}
n_\sigma = \frac{f_2 T_\text{exp}\langle \Gamma \rangle_T}{N_\text{tot}^{1/2}},
\label{eqn:modSignal}
\end{equation}
where $T_\text{exp}$ is the total exposure time, and $N_\text{tot}$ is the total number of events measured at the experiment. Assuming a constant background rate $R_b$ that scales linearly in both the exposure time and the target mass, the discovery potential increases with exposure time, even without background mitigation.

\subsection{Potential sensitivity to dark matter}

\begin{figure}[ht]
\centering
\includegraphics[width=\linewidth]{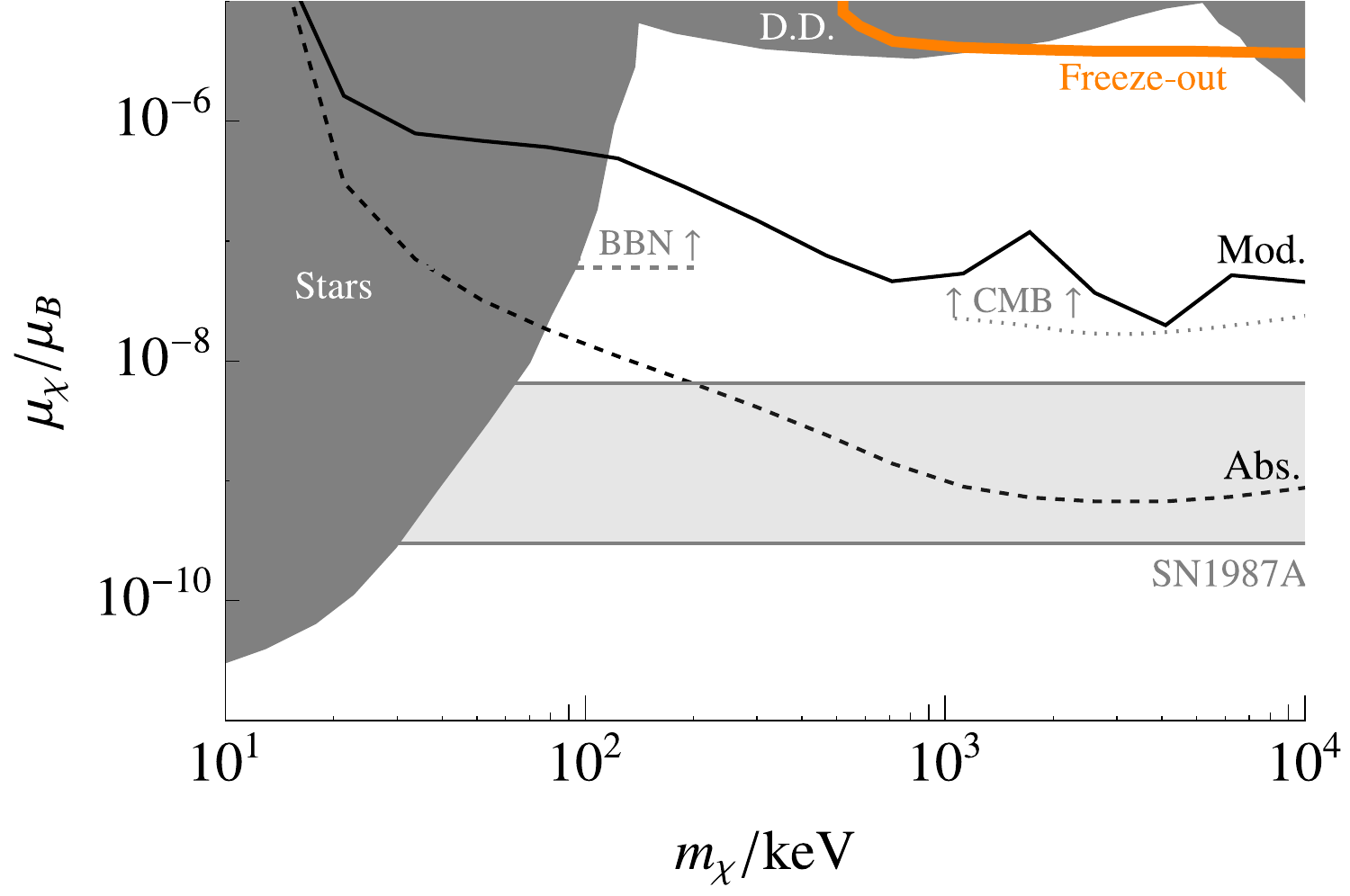}
\caption{The potential sensitivity of kg-year exposure of \ch{La2CuO4} to DM with a magnetic moment $\mu_\chi$ in units of the Bohr magneton $\mu_B = e/2m_e$. We take the electrons charge to be $e_V=e$. In solid black, we show the coupling at which the modulation of the DM signal is statistically signicant, as detailed in the text. In dotted black, we show the absolute limit of sensitivity, corresponding to three events with no background. Both projections assume a 25 meV threshold. We also show limits placed by stellar cooling~\cite{Chu:2019rok}, BBN, SN1987A, the CMB, and direction detection (DD)~\cite{Chu:2018qrm}; in orange, we show the parameters which reproduce the relic abundance through freeze-out.}
\label{fig:muChi}
\end{figure}

\begin{figure}[ht]
\centering
\includegraphics[width=\linewidth]{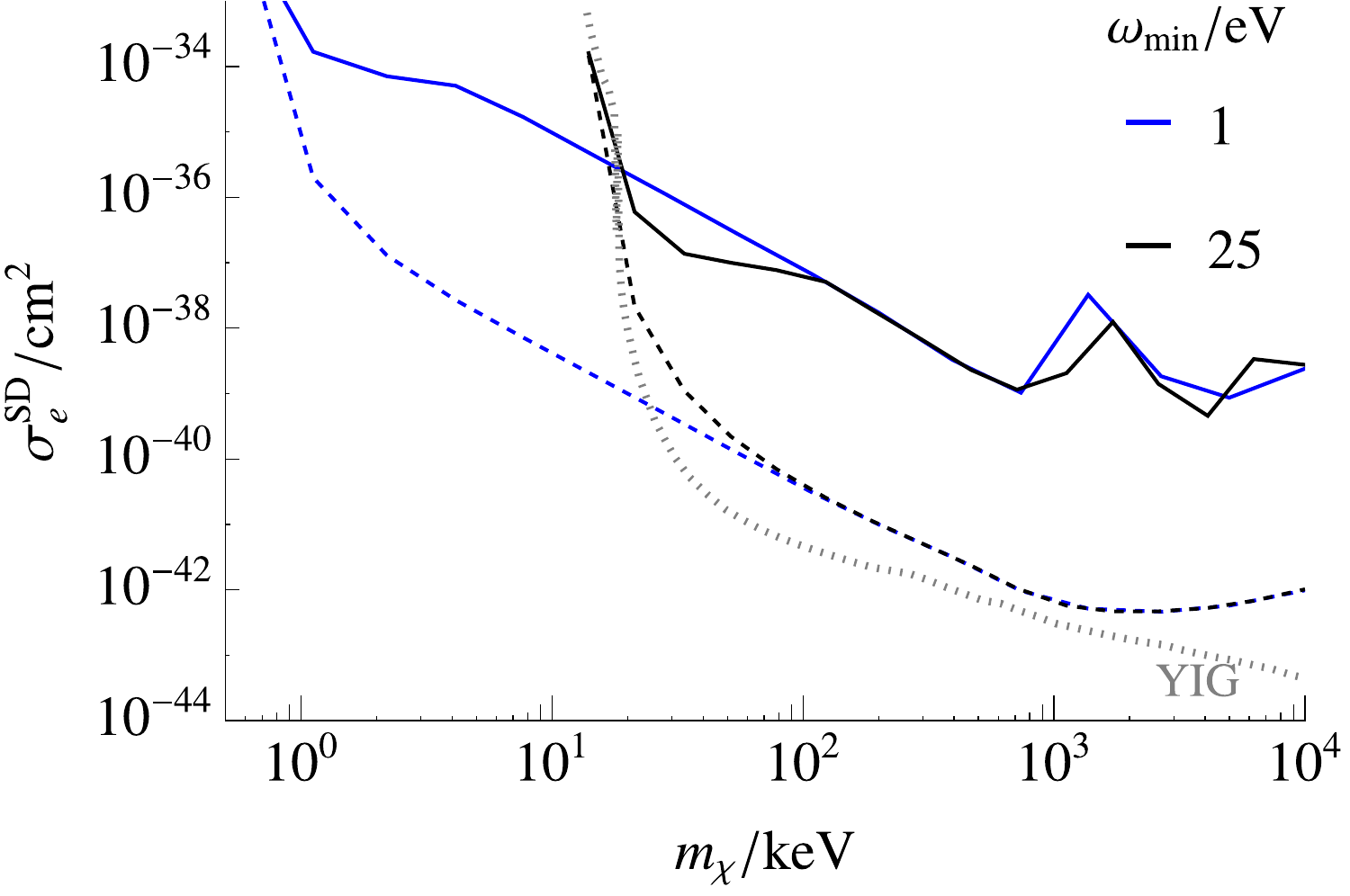}
\caption{Possible constraints on the standard SD interaction cross-section as a function of DM mass. Solid lines correspond to 3 sigma detection -- using Eqn. \eqref{eqn:modSignal} -- of a modulating signal over a constant signal for kg-year exposures. Dashed lines correspond to cross-sections that give 3 events per kilogram-year. We vary the energy threshold of $\omega_\mathrm{min}$ from 1 meV (blue) to 25 meV (black). We show for comparison the cross-section leading to 3 events/kg-year with a YIG target and a 25 meV threshold~\cite{trickleEffectiveFieldTheory2020a}. }
\label{fig:SDlimits}
\end{figure}

We now finally consider how these rates impact the potential sensitivity to dark matter.  We consider two statistical measures. The first corresponds to the 3 sigma detection of the DM daily modulation, given by Eq.~\eqref{eqn:modSignal}, assuming that all the DC events are signal dominated. This is an optimistic scenario in which we assume we have subtracted all non-DM induced background events, and are testing the hypothesis that the residual DC events are indeed DM-induced.

As a second measure of sensitivity, we also calculate the 95\% confidence limit at which any signal can be measured, again assuming zero background. This is a more aggressive measure than the previous modulation measure, which only takes into account the total rate of events. The total rate is straightforward to evaluate numerically, and its expression is given in Eq.~\eqref{eqn:rateFull} of App.~\ref{app:rate}

An important quantity that enters into evaluation of the projected sensitivity is the minimum energy threshold of a magnon, above which one can nominally detect its presence. Currently, single magnons are detectable by entangling the material with a qubit, which allows efficient readout of the Kittel magnon mode~\cite{lachance2020entanglement}, while to our knowledge the single-particle detection of other modes has yet to be demonstrated. 

One possible avenue would be through calorimetric detection of the magnons, as suggested in~\cite{trickleDetectingLight2020}. Such detection schemes are currently being developed for the SPICE experiment~\cite{Knapen:2017ekk, Hochberg:2015fth} as part of the TESSERACT collaboration, for which transition-edge sensors (TESs) are the tool of choice~\cite{Fink:2020noh, Romani:2024rfh}. The theoretical lower limit on the energy sensitivity of such a detector is twice the binding energy of the Cooper pair in the superconductor, which for an aluminium superconductor, for instance, corresponds to 7.2 meV. To reach this fundamental limit of energy resolution, detection efficiencies of order unity need to be achieved. As such, we choose to vary the threshold from between nominal values of 25 meV and 1 meV, with the latter value corresponding to the fundamental limit of a low-band-gap material.

For the model of DM interacting through a magnetic dipole, the existence of a light, electromagnetically coupled state can be probed regardless of its relic abundance. There are constraints from stellar cooling~\cite{Chu:2019rok} as well as supernova SN 1987A and cosmology~\cite{Chu:2018qrm}. Additional constraints exist from beam dump/collider experiments~\cite{Chu:2020ysb, Marocco:2020dqu}, although these rely on the MDM EFT holding up to scale $g m_\chi$, and so may be parametrically violated in a given UV model.

We show in Fig.~\ref{fig:muChi} the curves corresponding to the two measures just described. We see that the material we consider can be sensitive to DM in regions of parameter space not directly excluded by other probes. In particular, the modulating signal could be visible. Furthermore, we see that much of the interesting parameter space not probed by terrestrial/astrophysical measurements is at higher mass, at which point we find that the inclusion of the magnetic form factors becomes significant.

We also express limits in terms of the standard SD coupling. Here, we express the limits in terms of a ``reference cross-section" 
\begin{align}
    \bar{\sigma}_e &= |c|^2 m_e^{-2},
\end{align}
which converts bounds on $c$ to a cross-section with a free electron. The limits on this model are shown in Fig.~\ref{fig:SDlimits}, and compared with those from YIG. An important distinction in this case is that the use of quasi-2d magnets induces a daily modulation under the DM hypothesis, which would not occur in YIG (at least at long-wavelengths, where the dispersion relation is approximately isotropic). We further see that the use of AFMs allows reach at lower masses, down to around a keV, as previously noted in other AFMs~\cite{Esposito:2022bnu}. However, a given UV model of this DM candidate may face strong constraints from cosmology at such light masses, although a full exploration of this topic is beyond our current scope.

Lastly, it is important to highlight the stringent background constraints that the above projections assume. Such calorimetric detectors have observed large excesses of events at low energies~\cite{Fuss:2022fxe}, which are thought to arise from stresses in the target material~\cite{Anthony-Petersen:2022ujw, Anthony-Petersen:2024vdh}. This background presents a major hurdle to any search for a small number of dark matter-induced events, and it is an ongoing problem to mitigate or veto these events. A rigorous understanding of these and similar backgrounds is necessary to understand the limit of the proposed detection scheme. However, assuming the background events do not exhibit a daily modulation, one still has a sensitivity to the DM modulation, although at a reduced level depending on the rate of the DC-background according to Eq.~\eqref{eqn:modSignal}.


\section{Conclusions}

We have studied the scattering of keV-MeV mass DM that couples in electronic spins in magnetically ordered materials. In this mass regime, the relevant excitations are bosonic magnon degrees of freedom. Such DM candidates can excite a possibly measurable number of magnons in relevant regions of paramater space. At the lower end of the mass range, the energy of a magnon that can be exciting through scattering is in the meV range, below which calorimetric detection becomes challenging due to the finite binding energy of the superconducting pairs in a TES. 

At MeV DM masses, the momentum transfer is large enough to start resolving the electronic structure of the magnetic crystal. We have used the magnetic form factors to take this into account when matching to the magnon theory, and shown how this quantitatively changes the predicted cross-sections. Its inclusion is important to obtain an accurate prediction of the total DM signal rate, as well as its angular dependence. Far above this mass, the magnon theory breaks down,  and one expects to enter the single-particle scattering regime. 

We have also shown that quasi-2d ferromagnets allow for directional detection, even if the underlying interaction is isotropic. As we demonstrated, the daily modulation of the signal induced by the target anisotropy gives a vital handle on background discrimination.

A major outstanding question is the detection of the magnon excitations that may be produced in the DM scattering, as well as the associated backgrounds in this process. A robust understanding of these questions is needed in order to fully assess the potential of this method of dark matter detection. We hope that this work motivates a deeper understanding of the possibility of the limits of single magnon detection.

\section*{Acknowledgments}
We thank Omar Ashour, Sin\'ead Griffin, Simon Knapen, Alexander Millar, Mario Reig, Bethany Suter, Tanner Trickle, and Kevin Zhou for helpful discussions. We are especially grateful to Andrew Boothroyd for discussions related to magnetic scattering and low-dimensional magnets.  

\bibliographystyle{h-physrev}
\bibliography{references}

\appendix

\section{Magnons}
\label{app:magnons}
We give a concise introduction to the relevant properties of magnetic excitations. We focus on simple square-lattice ferromagnets and the linear spin-wave treatment of magnons. For more detailed information, see \cite{BoothroydAndrewT2020PoNS}.

Throughout this section, we work with the Heisenberg magnet \begin{equation}
\mathcal{H} = \sum_{\langle \alpha \beta \rangle} J_{\alpha \beta} \mathbf{S}_\alpha \cdot \mathbf{S}_\beta,
\label{appEqn:heisenberg}
\end{equation}
with a spin operator $\mathbf{S}_\alpha$ acting on each lattice site. A basis of states at a site is given by the eigenstates $\ket{m}_\alpha$ of $\sz$:
\begin{equation}
\sz \ket{m}_\alpha = m \ket{m}_\alpha
\end{equation}
for $m = -s,\ldots s$. We will also find it useful to define the spin raising and lowering operators 
\begin{equation}
S^\pm_\alpha = \sx \pm i \sy,
\end{equation}
in terms of which the above Hamiltonian is
\begin{equation}
\mathcal{H} = \sum_{\langle \alpha \beta \rangle} J_{\alpha \beta} \Big[\sz \sz[\beta] + \frac{1}{2}\left( \sP \sM[\beta] + \sM \sP[\beta] \right) \Big].
\end{equation}



\subsection{Ferromagnets} 
\label{app:ferromagnets}

\subsubsection{Finding the magnon Hamiltonian}
The sign of $J_{\alpha \beta}$ in Eq. \eqref{appEqn:heisenberg} is crucial in determining the ground state and the excitations above it. For $J_{\alpha \beta} = - \lvert J_{\alpha \beta}\rvert$, we have an exact ferromagnetic ground state, in which all the spins are aligned in the $z$-direction, say,
\begin{equation}
\ket{0}_{FM} = \prod_\alpha  \ket{s}_\alpha.
\end{equation}

Excitations around the ferromagnetic ground state may be studied by means of a Holstein-Primakoff transformation \cite{PhysRev.58.1098}. Defining the operators $\annihil$ by
\begin{equation}
\begin{split}
\sz &= s - \creation \annihil, \\
\sP &= \sqrt{2s\left( 1-\frac{\creation \annihil}{2s}  \right)} \annihil,
\end{split}
\label{appEqn:Holstein}
\end{equation}
we observe that the $\mathfrak{so}(3)$ commutation relations of $S^i_\alpha$ imply that the creation and annihilation operators obey the cannonical commutation relations for \textit{bosons}. The space of states spanned by the operators $S, a$ is in general different (by Fermi's exclusion principle), so one must impose the constraint that there are no more than $2s$ bosons at a site.

For small fluctuations around the ground state, we take the eigenvalues $n=s-m$ of the number operator
\begin{equation}
\creation \annihil \ket{n} = (s-m)\ket{n}
\end{equation}
to be small with respect to $s$. We may then express the Hamiltonian \eqref{appEqn:heisenberg} in terms of our new variables \eqref{appEqn:Holstein}, and linearise in $n/S$, restricting ourselves to work only with the low-lying states, thus forgetting about the constraint. The resulting Hamiltonian, to leading order, is 
\begin{equation}
\mathcal{H}_0 = \sum_\mathbf{q} \Omega(\mathbf{q}) a^\dag_\mathbf{q} a_\mathbf{q},
\label{appEqn:linearH}
\end{equation} 
where we have gone to momentum space
\begin{equation}
\annihil = N^{-1/2} \sum_\mathbf{q} e^{i \mathbf{q} \cdot \mathbf{x}_\alpha} a_\mathbf{q},
\end{equation}
 subtracted the zero-point energy,
and expressed the energy of each mode as $\Omega(\mathbf{q})$. Note that to this order in the expansion, the bosonic ladder operators are linearly related to the $S^\pm$:
\begin{equation}
\sP \simeq \sqrt{2s} \annihil.
\label{appEqn:linearisedOperators}
\end{equation}

 For a quasi-2d square lattice ferromagnet with lattice spacing $a$, in which all nearest-neighbours within a plane are coupled with the same strength $-J$, we have
\begin{equation} \Omega(\mathbf{q}) = 4 J s \left( \sin^2 \frac{q_x a}{2} + \sin^2 \frac{q_y a}{2} \right),
\end{equation}
where the sites are taken to be coupled in the $x-y$ plane. The excitations which carry this energy are called spin-waves, or magnons. Calculating the $xx$ and $xy$ correlations of the spin operator, we see  that they are $\pi/2$ out-of-phase, indicating that the spins are precessing coherently around the magnetisation direction.

\subsubsection{One-magnon correlations}
At low energies, the DM-target scattering will predominantly excite a single magnon. We thus take the excited state $\ket{f}$ to be a single magnon state, labelled by a momentum $\ket{\mathbf{q}}$. We now wish to calculate the matrix element of the spin operator in momentum space $\hatbf{S}(\mathbf{q})$ between the ground state $\ket{0}$ and this final state $\ket{\mathbf{q}}$
appearing in
\begin{equation}
S_{ij}(\omega,\mathbf{q})\equiv \frac{1}{V}  \bra{\mathbf{q}} \hat{S}_i(q) \ket{0} \bra{0} \hat{S}_j^\dagger(0) \ket{\mathbf{q}} \cdot \delta(\Omega(\mathbf{q}) - \omega).
\end{equation}
Observe that the linearised Hamiltonian in \eqref{appEqn:linearH} commutes with $\sz$, and that, by Eq. \eqref{appEqn:linearisedOperators}, the one-magnon state has $S^z$ eigenvalue $s-1$. Hence $S_{zi} = S_{xy}= 0$ and $S_{xx} = S_{yy} \equiv \mathcal{S}_\mathrm{FM}$ (after symmetrising the spatial indices). A straightforward calculation gives the only non-zero component of the matrix at zero-temperature:
\begin{equation}
\mathcal{S}_\mathrm{FM} = n_S \delta(\omega - \Omega(\mathbf{q})),
\end{equation}
where $n_S$ is the effective spin density per volume of the ferromagnet. 

\subsection{Antiferromagnets}
\label{app:AFMs}
The model of antiferromagnets we consider is again of interacting lattice spins with a Heisenberg hamiltonian
\begin{equation}
\mathcal{H} = \sum_{\langle \alpha \beta \rangle} J_{\alpha \beta} \mathbf{S}_\alpha \cdot \mathbf{S}_\beta,
\label{appEqn:heisenbergAgain}
\end{equation}
but we now have $J_{\alpha \beta} = \lvert J_{\alpha \beta} \rvert$, meaning that the classical ground state has neighbouring spins anti-aligned. Such systems have been much studied, see for instance~\cite{AFMreview} for a review. This ground state, in which the magnon is the Ne\'{e}l phase is not the true quantum ground state of the system, although in the semiclassical regime it gives accurate results. 

Such a collinear AFM is equivalent to a sum of two sublattices, each with their own creation and annihilation operators. Labelling these sublattices $A$ and $B$, we apply individual Holstein-Primakoff transformations to them, which to leading order are
\begin{align}
        S_{A, \alpha}^z &= S - a^\dag_\alpha a_\alpha, & S_{B, \alpha}^z &= -S + b^\dag_\alpha b_\alpha, \\
        S_{A \alpha} ^+ &\simeq \sqrt{2S}a_\alpha, & S_{B \alpha} ^+ &\simeq \sqrt{2S}b_\alpha^\dag.
\end{align}
In contrast to the previous FM case, we now have two sets of bosonic operators $a_\alpha$ and $b_\alpha$. In the $1/S$ expansion, the leading non-constant term is bilinear in these operators, and the Hamiltonian may generally be written in momentum space as

\begin{equation}
\mathcal{H} = -S^2 \mathcal{H}_0 + S\sum_\mathbf{q} \mathbf{y}^\dag_\mathbf{q} M_\mathbf{q} \mathbf{y}_\mathbf{q} + \mathcal{O}(S^{0}),
\end{equation}
where the vector operator $\mathbf{y}_\mathbf{q}$ is
\begin{equation}
    \mathbf{y}_\mathbf{q} = \begin{pmatrix}
        a_\mathbf{q} \\ b_\mathbf{q} \\ a^\dag_\mathbf{-q} \\ b^\dag_\mathbf{-q}
    \end{pmatrix},
\end{equation}
and $M_\mathbf{q}$ is a matrix whose components depend on the particular couplings $J_{\alpha \beta}$ of the original Heisenberg hamiltonian. 

In our analysis, we will only make use of isotropic AFMs, such that the matrix is given by
\begin{equation}
M_\mathbf{q} = 
  J(0)  \begin{pmatrix}
        1 & 0 & 0 & m_\mathbf{q} \\
        0 & 1 & m_\mathbf{q} & 0 \\
        0 & m_\mathbf{q} & 1 & 0 \\
        m_\mathbf{q} & 0 & 0 & 1
    \end{pmatrix},
\end{equation}
where the matrix element $m_\mathbf{q}$ is
\begin{equation}
    \begin{split}
        m_\mathbf{q} &= \frac{ \cos(q_x a) + \cos(q_y a)}{2}.
    \end{split}
\end{equation}
To obtain the propagating magnon modes, we must diagonalise the mass matrix; we find a pair of degenerate modes with frequency 
\begin{align}
\Omega(\mathbf{q}) &= S J(0) \sqrt{1 - m_\mathbf{q}^2}. 
\end{align}

Summing over both of the degenerate branches, the one magnon contribution to the low-temperature limit of the spin structure function is then
\begin{equation}
    \begin{split}
        S_{xx}(\mathbf{q}) &= S_{yy}(\mathbf{q}) = n_S J(0) \frac{1 - m_\mathbf{q}}{\Omega(\mathbf{q})} \\
        &= n_S  \sqrt{\frac{1 - m_\mathbf{q}}{1 + m_\mathbf{q}}} \\
        &= n_S \, \sqrt{\frac{2 - \cos q_x a - \cos q_y a}{2 + \cos q_x a + \cos q_y a}},
    \end{split} 
\end{equation}
with the other components vanishing.

\subsection{Quasi-1D AFMs}

In this case, we again have a pair of degenerate magnon modes, each with frequency 
\begin{equation}
    \Omega(\mathbf{q}) = 2JS \Big| \sin q_z a \Big|,
\end{equation}
and 
\begin{equation}
    S_{xx} = S_{yy} = 2n_S\tan\frac{q_z a}{2}\,\delta\left(\omega-\Omega(\mathbf{q})\right)
\end{equation}


\section{Magnetic form factors}
\label{app:formFactor}

For the Cu$^{2+}$ magnetic ions of interest here, there is a single unpaired electron in a $3d_{x^2-y^2}$ orbital \cite{PhysRevB.48.13817}, and so the form factor is \cite{BoothroydAndrewT2020PoNS}
\begin{align}
\begin{split}
f(\mathbf{q}) &= \langle j_0 \rangle + \frac{5}{7}\left( 3 \cos^2 \theta -1\right) \langle j_2 \rangle \\
&+ \frac{3}{56}\left(- 30 \cos^2 \theta  + 35 \cos^4 \theta + 35 \sin^2 \theta \cos 4 \phi  \right) \langle j_4 \rangle
\label{eqn:3dFormFactor}
\end{split}
\end{align}
where $\theta$ and $\phi$ are the polar and azimuthal angles, respectively, with the $x-y$ plane of the orbital taken along the equator; the functions $\langle j_n \rangle (q)$ are the weighting of the $n$th Bessel function over the radial wavefunction of the electron arising from an expansion in $qR$, with $R$ the ionic radius. This form factor is illustrated in Fig. \ref{fig:anistropicForm}. A convenient phenomenological parametrisation is given by
\begin{equation}
\langle j_n \rangle (4\pi q) = q^2_n \left( \sum_{i=1}^{4} A_{n,i}e^{-a_{n,i}q^2} \right),
\end{equation}
where the prefactor is $q^2_0 =1$ or $q^2_m = q^2, n = 2,4,6$; the constants $A_{n,i}, a_{n,i}$ are taken from  The Cambridge Crystallographic Subroutine Library.


\section{Relativistic models}

\subsection{Magnetic dipole moments}
\label{app:MDM}

Consider a dark photon $V_\mu$ with a minimal coupling to electrons with charge $e_V$, and which couples through a magnetic dipole interaction to a dark state $\chi$
\begin{align}
    \mathcal{L}_\mathrm{int} = e_V V_\mu \, \bar{e} \gamma^\mu e + \frac{1}{2} \mu_\chi \bar{\chi} \sigma^{\mu\nu} \chi V_{\mu\nu},
\end{align}
where $\sigma^{\mu \nu} = \frac{i}{2}[\gamma^\mu, \gamma^\nu]$.
In the non-relativistic limit, the dark state $\chi$ couples to the dark magnetic field $\mathbf{B}_V$ 
\begin{align}
    H_\mathrm{int} \supset - 2 \mu_\chi \mathbf{B}_V \cdot \mathbf{S}_\chi
\end{align}
via its spin $\mathbf{S}_\chi = \boldsymbol{\sigma}_\chi/2$, where $\boldsymbol{\sigma}_\chi$ are the Pauli matrices acting on the $\chi$ subspace. Meanwhile, an electron's spin $\mathbf{S}_e = \boldsymbol{\sigma}_e/2$ sources a dark magnetic field, which for a massless dark photon is
\begin{align}
    \mathbf{B}_V(\mathbf{x}) = - \frac{e_V}{4\pi m_e} \nabla \times \left(\frac{\mathbf{S}_e\times \mathbf{x}}{x^3} \right).
\end{align}
The interaction potential in momentum space is then given by
\begin{align}
    V(\mathbf{q}) = - 2\frac{e_V \mu_\chi}{m_e} \mathbf{S}_\chi \cdot \Pi_\perp(\hat{q}) \cdot \mathbf{S}_e,
\end{align}
where $\Pi_\perp(\hat{q})$ projects onto the subspace perpendicular to the direction of momentum transfer:
\begin{align}
    \Pi_\perp(\hat{q})_{ij} := \delta_{ij} - \hat{q}_i \hat{q}_j. 
\end{align}
The DM-spin-averaged/summed interaction potential is 
\begin{align}
    V_{ij}(\mathbf{q}) = 2 \left( \frac{e_V \mu_\chi}{m_e}\right)^2 \Pi_{\perp,ij}(\hat{q}),
\end{align}
where we have made use of Eq. \eqref{eqn:spinspin}. For simplicity, we will take $e_V = e$ when expressing bounds on an MDM.

\subsection{Heavy axial-vector}
\label{app:axialVector}
The bounds we place may also be mapped on to a model where an axial-vector gauge boson $V_\mu$ of mass $m_V$ couples to the axial-vector DM and electron currents:
\begin{align}
    \mathcal{L}_\mathrm{int} = g_V V_\mu \, \bar{\chi}\gamma^\mu \gamma^5 \chi + g_e V_\mu \bar{e} \gamma^\mu \gamma^5 e.
\end{align}
If $m_V \gg 10^{-3} m_\chi$,  the gauge boson mass is much larger than the typical scattering momentum transfer, and the Fourier transform of the interaction Hamiltonian is
\begin{align}
    V(\mathbf{q}) = \frac{4 g_\chi g_e}{m_V^2} \mathbf{S}_\chi \cdot \mathbf{S}_e.
\end{align}
The spin-summed interaction potential is then
\begin{align}
    V_{ij} = \left( \frac{4 g_\chi g_e}{m_V^2} \right)^2 \delta_{ij}.
\end{align}

\subsection{Spin sums}
In scattering cross-sections, we take the initial DM state to be unpolarised, and we sum over final DM spin-states. In doing so, we make use of the identity
\begin{align}
    \frac{1}{2} \sum_{S_\chi, S'_\chi}  S_\chi^i S_\chi^{'j} = \frac{1}{2} \delta^{ij},
    \label{eqn:spinSum}
\end{align}
which holds for the spin-1/2 DM with $\mathbf{S}_\chi = \boldsymbol{\sigma}/2$.

\section{Rate and phase space integrals}
\label{app:rate}

We now have all the ingredients necessary to calculate the DM-scattering rate as a function of DM velocity. To obtain a total expected rate, we must average over the dark matter velocity distribution $f_\chi(\mathbf{v})$. We take the standard form of a Maxwell-Boltzmann distribution truncated at the sphere of radius $v_\text{esc}$ in the galactic frame, i.e. in the laboratory frame we have
\begin{equation}
f_\chi(\mathbf{v}) = \frac{1}{N_0} \exp{\Big[-(\mathbf{v}+\mathbf{v_e})^2/v_0^2\Big]} \Theta(v_\text{esc}- \lvert \mathbf{v}+\mathbf{v_e} \rvert),
\end{equation}
where the normalisation $N_0$ is
\begin{equation}
N_0 = \pi v_0^2 \left( \pi^{1/2} v_0\, \text{Erf}(v_\text{esc}/v_0) - 2 v_\text{esc}e^{-v_{\text{esc}}^2/v_0^2}  \right).
\end{equation}
The numerical values we use are $v_0 = \SI{230}{km/s}$, $v_\text{esc} = \SI{600}{km/s}$, and $v_e = \SI{240}{km/s}$. The average rate is in general given by
\begin{equation}
\Gamma = \int d^3 \mathbf{v}\, f_\chi(\mathbf{v}) \Gamma(\mathbf{v}),
\end{equation}
which involves six integrals, where $\Gamma(\mathbf{v})$ is given by Eq.~\eqref{eq:rate}. This is related to the average rate per target mass by
\begin{equation}
R = \rho_T^{-1}\frac{\rho_\chi}{m_\chi}\Gamma.
\end{equation}

Carrying out the angular velocity integrals in the galactic frame, where velocities are described by $\mathbf{v}' = \mathbf{v} - \mathbf{v}_\oplus$, we find
\begin{align}
    \Gamma = &2\pi \times \int_0^{v_\mathrm{esc}} dv'  \int \frac{d^3\mathbf{q}}{(2\pi)^3} \, f_\chi(v') \, \frac{v'}{q} \nonumber \\
    &\times V_{ij}(\mathbf{q}) S^{ij}(\mathbf{q})  \, \Theta\big(v' - v_\mathrm{min}'(\mathbf{q})\big),
\end{align}
where the factor of $2\pi$ arises from the azimuthal integral, and we have used the Dirac delta function to integrate over the polar angle, and 
\begin{align}
    v_\mathrm{min}'(\mathbf{q}) \equiv \frac{q}{2m_\chi} + \frac{\Omega_\mathbf{q} + \mathbf{v}_\oplus \cdot \mathbf{q}}{q}.
\end{align}
Carrying out the final integral over $v'$, we find
\begin{align}
    \Gamma =  &\frac{v_0^2}{N_0} \, \pi \times \int \frac{d^3\mathbf{q}}{(2\pi)^3} \frac{\Theta\big(v'_\mathrm{min}(\mathbf{q})\big)}{q} \nonumber \\
    \times &V_{ij}(\mathbf{q}) S^{ij}(\mathbf{q})\left( e^{-v'_\text{min}(\mathbf{q})^2/v_0^2} - e^{-v ^2_\text{esc}/v_0^2} \right) .
    \label{eqn:rateFull}
\end{align}

\section{Domains and neutron scattering}
\label{sec:nonPert}
The results so far have strictly applied only to one-magnon excitations within the framework of linear spin-wave theory. In this section, we discuss calibration to experimental data in the presence of magnetic domains, in order to do away with these assumptions.  Our method is analogous to the proposal of \cite{hochbergDeterminingDarkMatter2021, knapenDarkMatterelectron2021}, where it was pointed out that one can directly measure the part of the correlation function relevant for DM scattering from coupling to the electron number density. We also discuss how magnetic domains that enable experimental calibration also wash away directionality in otherwise isotropic materials.

We begin with the observation that the cross section for magnetic neutron scattering from some initial momentum $k_i$ to a final momentum $k_f$ is 
\begin{equation}
\frac{d^3 \sigma}{d \Omega d E_f} = \frac{k_f}{k_i} \left( \frac{\gamma r_0}{2\mu_B} \right)^2  |f(\mathbf{q})|^2(\delta_{ij} - \hat{q}_i \hat{q}_j) \tilde{S}_{i j} ( \omega, \mathbf{q}).
\label{eqn:neutronScattering}
\end{equation}
Since neutron scattering proceeds via a magnetic dipole interaction, only the piece of the spin-spin correlation orthogonal to the direction of momentum transfer contributes. This projection onto the orthogonal subspace  naively precludes a direct comparison between neutron- and DM-scattering.

One way forward is to note that we may in fact rotate away the $\hat{q}$-dependence in the projector if the crystal is made up of many domains, with the direction of a particular domain's magnetisation being isotropically distributed \cite{lorenzanaSumRules2005}. This is indeed the case in many crystals of interest, in particular the ones studied here.

\subsection{Domains}
\label{app:domains}

Since we are interested in the physics of multi-domain systems, let us rewrite the spin correlator in a way that brings out the structure of the domains. The relevant function is
\begin{equation}
\tilde{S}_{ij}(\omega, \mathbf{q}) = \int \, dt e^{-i \omega t} \sum_{\alpha \beta} e^{i \mathbf{q}\cdot (\mathbf{X}_\alpha-\mathbf{X}_\beta)}\langle S_{\alpha,i}(t) S_{\beta,j}(0) \rangle,
\end{equation}
where we place a tilde to emphasise that the matrix elements are at the level of the lattice, and so omit form factors, etc; recall that the sum over $\alpha, \beta$ indicates a sum over all sites on the lattice. Note that only the symmetric (in $i,j$) part of this contributes to scattering processes. We may split this sum into a sum over domains $D$ and a further sum over sites within this domain, which we denote
\begin{equation}
\sum_\alpha = \sum_D \sum_{\alpha \in D}.
\end{equation}
We now assume that spins in different domains are uncorrelated. Physically, this means that the magnons cannot propagate between domains -- they are scattered by the domain walls and the coherent wave is destroyed. Thus, the correlator is diagonal in $D$-space.

To evaluate the spin-correlation function, we have implicitly been assuming a fiducial co-ordinate system in which we quantise the spins along a preferred $z$-axis. In a multi-domain system, it is convenient to have a local co-ordinate system where the magnetisation direction is aligned with the $z$-direction. We will denote quantities evaluated in this system with a prime $'$, and we introduce a domain-dependent rotation matrix that takes us from the local co-ordinate system to the global one:
\begin{equation}
\langle S_{\alpha}(t) S_{\beta}(0) \rangle = R \cdot  \langle S'_{\alpha}(t) S'_{\beta}(0) \rangle \cdot R^T.
\end{equation}

In terms of the local system, the projection of $\tilde{S}$ is
\begin{align}
\Pi_{\perp,ij}(\hat{q})\tilde{S}_{ij}(\omega, \mathbf{q}) &=  \sum_D (\delta_{ij} - \hat{q}^D_{i} \hat{q}^D_j)  \int \, dt e^{-i \omega t} \\
&\times \sum_{\alpha,\beta \in D} e^{i \mathbf{q}\cdot (\mathbf{X}_\alpha-\mathbf{X}_\beta)} \langle S'_{\alpha,i}(t) S'_{\beta,j}(0) \rangle,
\label{eqn:domainSum}
\end{align} 
where $\hat{q}^D$ is the rotated momentum unit vector: $\hat{q}^D = R^D \cdot \hat{q}$. To evaluate this sum, a final assumption must now be made: let all the domains be approximately equal, with the magnetisation direction being on average isotropic, i.e.  $[\mathcal{H}, \sum_\alpha S_\alpha^i] = 0.$. This means we can translate a sum over the domain-specific rotations into an isotropic integral over rotation matrices. Explicitly, we parametrise the rotations $R^D$ in terms of a rotation axis $\hat{n}$, specified by polar co-ordinates $(\theta, \phi)$ and an angle $\gamma$, for which the normalised, isotropic Haar measure is \cite{10.2307/2333716} $d\Omega = \frac{1}{2\pi^2}\sin^2 \frac{\gamma}{2} d\gamma\, d\cos \theta\, d \phi$ and we have
$\hat{q}^D_i = \left(\cos \gamma \delta_{ij}  +\sin \gamma\, \epsilon_{ikj} \hat{n}_k  +(1-\cos \gamma) \hat{n}_i \hat{n}_j  \right) \hat{q}_j.
$
As is evident from Eq. \eqref{eqn:domainSum}, to evaluate the projected correlator we only require knowledge of
\begin{equation}
\int d\Omega \left(R(\Omega) \cdot \hat{q}\right)_i\left(R(\Omega) \cdot \hat{q}\right)_j = \frac{1}{3} \delta_{ij},
\end{equation}
making use of $\int d\Omega \, (\hat{q}\cdot R \cdot \hat{q})^2 = 1/3 $ and $\int d\Omega \, \lvert R \cdot \hat{q} \rvert^2 = 1$.
This demonstrates that the isotropic distribution of domains simply effects a calculable rescaling of the scattering potential, allowing one to determine $\tilde{S}_{ii}$ directly from neutron scattering data. Similar observations exist in the neutron scattering literature, e.g. \cite{lovesey,lorenzanaSumRules2005}. 



Finally, note that even though the magnetisation directions are isotropically distributed, there can still be a daily modulation of a DM signal. The couplings of the underlying lattice nonetheless break the isotropy. One may think of the spin isotropy as being an internal symmetry of the system, meaning that the anisotropy of the dispersion relation and form factors in real space still contribute, even while the anisotropy of the scattering potential is averaged out.

\end{document}